# Wideband photonic blind source separation with optical pulse sampling

TAICHU SHI,[1] YANG QI,[1] WEIPENG ZHANG,[2] PAUL PRUCNAL,[2] JIE LI,[1] AND BEN WU[1,*]

[1]*Department of Electrical and Computer Engineering, Rowan University, Glassboro, New Jersey 08028 USA*
[2]*Lightwave Communications Laboratory, Department of Electrical Engineering, Princeton University, Princeton, NJ 08544, USA*
*\*wub@rowan.edu*

**Abstract:** We propose and experimentally demonstrate an optical pulse sampling method for photonic blind source separation. The photonic system processes and separates wideband signals based on the statistical information of the mixed signals and thus the sampling frequency can be orders of magnitude lower than the bandwidth of the signals. The ultra-fast optical pulse functions as a tweezer that collects samples of the signals at very low sampling rates, and each sample is short enough to maintain the statistical properties of the signals. The low sampling frequency reduces the workloads of the analog to digital conversion and digital signal processing systems. In the meantime, the short pulse sampling maintains the accuracy of the sampled signals, so the statistical properties of the undersampling signals are the same as the statistical properties of the original signals. With the optical pulses generated from a mode-locked laser, the optical pulse sampling system is able to process and separate mixed signals with bandwidth over 100GHz and achieves a dynamic range of 30dB.

## 1. Introduction

The substantially increased use of radio frequency (RF) spectrum generates wireless interferences. With the emerging as well as the evolving technologies in wireless communication systems that seek for higher frequencies for large bandwidth, the congestion of RF spectrum does not only exist under 6GHz (sub-6), but also appears in higher frequencies. Blind source separation (BSS) has been proved to be an effective way to manage such interferences [1,2]. By using multiple-input and multiple-output (MIMO) receivers, the BSS system separates the signal of interest (SOI) from interferences without pre-known information of the signals [3,4]. Traditional BSS methods separates the mixed signals with digital signal processing (DSP). The digital methods achieve MHz to GHz bandwidth and require the pre-known carrier frequencies, so the bandpass signals can be converted to baseband signals for analog to digital conversion (ADC) and BSS. For signals with bandwidths beyond GHz range, the high data rates generate considerable amount of workload to both the ADC and DSP, and thus it is challenging to achieve real-time signal separation of wideband signals. More importantly, for passive users, such as radio telescope, the carrier frequencies of SOI are unknown and could range from a few MHz to 100GHz [5,6]. Therefore, a BSS system that is capable of processing ultra-wideband signals is needed for both the growing needs of bandwidth from active users and the requirement of passive users.

Photonic signal processing provides a bandwidth independent method for BSS [7]. By modulating the RF signals on optical carriers, THz bandwidth is available at optical communication bands (193THz) [8]. Photonic BSS has been recently implemented with photonic integrated circuit [9,10]. One of the key advantages of photonic BSS is that the signals are separated in an analog way without ADC. Only a very small portion of the signals needs to be digitized to measure the statistical properties of the mixed signals and thus calculate the de-mixing matrix. The sampling frequencies can be orders of magnitude smaller than the bandwidths of the mixed signals, which greatly reduces the workloads of ADC and DSP. To sample the ultra-wideband signals (100GHz) at sub-Nyquist sampling rate, the major challenge is the sampling time, and it is impractical for the sampling circuit to achieve pico-second or femto-second sampling time with MHz or kHz sampling rate.

In this paper, we take advantage of the undersampling properties of photonic BSS for wideband RF signal processing and modulate the RF signals on optical pulses that pre-sample the signals before ADC. By using a mode-locked laser, the optical pulse width is in the order of 100fs. The ultra-narrow optical pulse function as a tweezer that collects samples of the mixed signals. Only a tiny potion of the mixed signals ($1/10^2$ to $1/10^5$) is collected, which greatly reduces the workload of ADC and DSP, while each collected sample is short and thus accurate enough to represent the statistical information of the mixed signals. The DSP system calculates the de-mixing matrix based on the statistical information and controls the photonic circuit to separate the SOI from interferences with the de-mixing matrix.

The relationship between the optical pulse sampling and photonic BSS is similar to the relationship between the biopsy sample preparation and curative surgery. The biopsy (pulse sampling) collects samples for analysis (DSP) and curative surgery (BSS) is performed based on the analysis. Traditional digital BSS method digitizes the mixed signals at Nyquist sampling rate and works in a similar way as cutting the whole tissue off for sample analysis, which is not efficient and introduce extra workloads to both sample preparation and analysis. The ultra-fast optical pulse acts as a tweezer for sample preparation. With a small portion of the mixed signals sampled for statistical analysis, the workloads of both sample preparation (ADC) and analysis (DSP) are reduced for real-time and ultra-wideband signal processing.

## 2. Principle and experimental setup

This section discusses the principle and the experimental setup of photonic BSS system with optical pulse sampling. Section 2.1 discusses the principle of photonic BSS, and shows that the system can separate the mixed signals with sampling frequency lower than the Nyquist sampling rate. Section 2.2 discusses the principle of optical pulse sampling.

### 2.1 Principle of blind source separation with undersampling

Fig. 1 shows the schematic diagram of the photonic circuit that implement BSS algorithm. The photonic BSS system processes mixed signals from a MIMO receiver. The mixed signals are represented by:

$$\boldsymbol{X} = \boldsymbol{AS} \text{ or, } \begin{bmatrix} x_1 \\ x_2 \end{bmatrix} = \begin{bmatrix} a_{11} & a_{12} \\ a_{21} & a_{22} \end{bmatrix} \begin{bmatrix} s_{soi} \\ s_{int} \end{bmatrix} \quad (1)$$

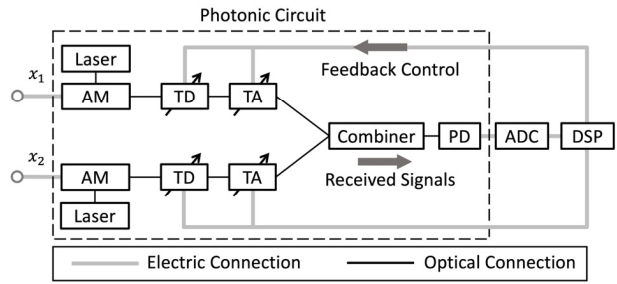

Fig. 1 Schematic diagram of the photonic BSS system (AM: optical amplitude modulator, TD: tunable optical delay, TA: tunable optical attenuator, PD: photodetector, ADC: analog to digital conversion, DSP: digital signal processing).

$A$ is the mixing matrix. $x_1$ and $x_2$ are received mixed signals from MIMO antennas. To demonstrate the principle of photonic BSS, a 16QAM is used as the SOI $s_{soi}$, and a Gaussian noise is used as the interference $s_{int}$. Both SOI and interference are baseband signals with the same bandwidth. Fig. 2 (a) is the scatter plot that shows the real parts of the mixed signals $x_1$ and $x_2$. The four straight lines match with the pattern of 16QAM, which corresponds to the projection of 16QAM on the real axis. Within each line, the density of the dots is Gaussian distributed, which represents the pattern of the interference. Separating SOI and interference is to find the de-mixing matrix $A^{-1}$, which includes two steps, principal component analysis (PCA) and independent component analysis (ICA) [7]:

$$A^{-1} = VU\Sigma U^{-1} \quad (2)$$

Where $U\Sigma U^{-1}$ represents the PCA, and $V$ represents the ICA. To perform PCA, the 2nd order moments of the mixed signals are measured with different weights applied to the signals. By using the tunable attenuations in the photonic circuit (Fig. 1), the following weights are added to the mixed signals:

$$x_{PCA} = \cos(\theta)x_1 + \sin(\theta)x_2 \quad (3)$$

Where $\theta$ is defined in Fig. 2 (a). Based on Equation 3, the 2nd order moment of $x_{PCA}$ is a function of $\theta$ (red curve in Fig. 2 (a)):

$$E(x_{PCA}^2) = q_1 + q_2 \cos[2(\theta - \theta_0)] \quad (4)$$

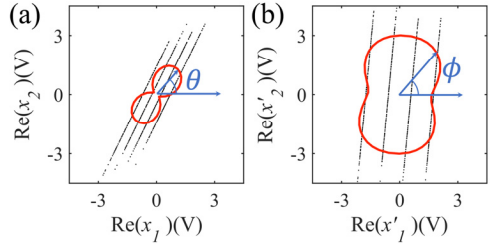

Fig. 2. (a) Real parts of the mixed signals, the red curve shows the 2nd order moments of the mixed signals; (b) Real parts of the mixed signals after PCA, the red curve shows the 4th order moments the mixed signals after PCA.

Where $E(x_{PCA}^2)$ is the time average of $x_{PCA}^2$. $\theta_0$ shows the direction of first principal component. $q_1 + q_2$ is the magnitude of the first principle component, and $q_1 - q_2$ is the magnitude of the second principal component. With three random $\theta$ values, the three unknown parameters $q_1$, $q_2$, and $\theta_0$ are determined. The PCA is performed with $X' = [x'_1, x'_2] = U\Sigma U^{-1}X$, where:

$$U = \begin{bmatrix} \cos(\theta_0) & -\sin(\theta_0) \\ \sin(\theta_0) & \cos(\theta_0) \end{bmatrix} \quad (5)$$

$$\Sigma = \begin{bmatrix} 1 & 0 \\ 0 & \sqrt{\frac{q_1+q_2}{q_1-q_2}} \end{bmatrix} \quad (6)$$

The PCA normalizes the 2nd order moments and whitens the mixed signals. To perform ICA, the 4th order moments (kurtosis) of the whitened signals are measured (red curve in Fig. 2 (b)) with new weights applied:

$$x_{ICA} = \cos(\phi)x'_1 + \sin(\phi)x'_2 \quad (7)$$

$$E(x_{ICA}^4) = p_1 + p_2\cos[2(\phi - \phi_0)] + p_3 \cos[4(\phi - \phi_0)] \quad (8)$$

Where $\phi$ is defined in Fig. 2 (b). With four random $\phi$ values, the unknown parameters $p_1$, $p_2$, $p_3$, and $\phi_0$ are determined, and ICA can be performed with $\mathbf{S} = \mathbf{VX'}$, where:

$$\mathbf{V} = \begin{bmatrix} \cos(\phi_0) & -\sin(\phi_0) \\ \sin(\phi_0) & \cos(\phi_0) \end{bmatrix} \quad (9)$$

The de-mixing matrix is solved as $\mathbf{A^{-1}} = \mathbf{V U \Sigma U^{-1}}$. With the de-mixing matrix, the SOI is separated from the interference. The tunable delays and attenuators in Fig. 1 control the complex weights for $x_1$ and $x_2$ and the circuits within the dashed lines represent half of the de-mixing matrix $\mathbf{A^{-1}}$, which recovers the SOI. Based on Equation 4 and Equation 8, only statistical properties of the mixed signals are needed to find the de-mixing matrix $\mathbf{A^{-1}}$. To obtain the statistical properties, which are 2nd and 4th order moments, a sampling rate orders of magnitude lower than the Nyquist sampling rate can be applied. The undersampling rate reduces the workload of DSP, and the proposed system is able to process mixed signals with bandwidth that is beyond ADC speed limit.

### 2.2 Optical pulse sampling

When the sampling rate is orders of magnitude smaller than the Nyquist frequency, the duty cycle $\tau/T_s$ also needs to be small enough, so the sampled signals can accurately represent the statistical information of the original signals. If the time duration $\tau$ of each sample is longer than the bit length of SOI, the sampled signals do not accurately represent the statistical information of the mixed signals. This can be explained by $E(x_1^2) \neq E(x_1)^2$, or in another word, when measuring the 2nd order moments, the time average of the square does not equal to the square of the time average. The sine function in Fig. 3 (a) shows the maximum frequency of the mixed signal $x_1$. In a photonic BSS system, the sampling frequency ($1/T_s$) can be less than the frequency of the sine function, while the duration of each sample has to be small enough ($\tau_1$ in Fig. 3 (a)). If the duration of each sample ($\tau_2$ in Fig. 3 (a)) is comparable to $T_s$, each sample is an average of the changing signals, and the sampled signals cannot represent the distribution of the original signals.

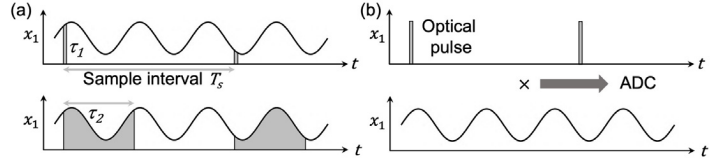

Fig. 3 (a) Comparison of short sampling time $\tau_1$ and long sampling time $\tau_2$ (b) Sample the signals with optical pulses

To obtain extremely small duty cycle ($10^{-4}$ to $10^{-8}$) for undersampling, the RF signals are modulated on a pulse laser that pre-samples the signals before ADC. The output of the optical modulator is the product of the periodic laser pulses and the RF signals (Fig. 3 (b)). The sampling time is determined by the laser pulse width, so even ADC with long sampling time $\tau_2$ is used afterwards; the sampled signals still represent accurate statistical information from the short sample duration $\tau_1$.

### 3. Experimental results and analysis

In this section, optical pulse sampling for BSS is demonstrated with experimental results. The optical pulses are generated with both optical intensity modulator (section 3.1) and mode-locked laser (section 3.2). With optical intensity modulator, optical pulses sample RF signals with sub-6GHz frequencies. Such system can be used to separate multiple wireless systems that coexist at the same RF frequencies, including 4G LTE, weather radar, radio telescope, etc. With the mode-locked laser, the optical pulses sample signals with 100GHz bandwidth and beyond. The pulse width of the mode-locked laser is in the order of 100fs, which functions as a tweezer that collect samples from ultra-wideband signals and obtain accurate statistical information. Mixed signals with mmWave frequencies or even THz frequency can be sampled with mode-locked laser pulses, which meets with the growing needs for 5G communication network [11,12], mmWave radar [13,14], and THz communication [15,16].

### 3.1 Demonstration of BSS with optical pulse sampling

In this section, we experimentally demonstrated that signals sampled with optical pulses accurately represent the statical properties. With 2$^{nd}$ and 4$^{th}$ order moments measured from the sampled signals, de-mixing matrix is solved. Fig. 4 shows the basic concept of optical sampling with a binary signal. The sampled signals (Fig. 4 (d)) is a product between the laser pulse sequence (Figs. 4 (a) and (b)) and the unsampled signals (Fig. 4 (c)). In this experiment, the laser pulse width is 5ns, and the signal bandwidth is 200MHz.

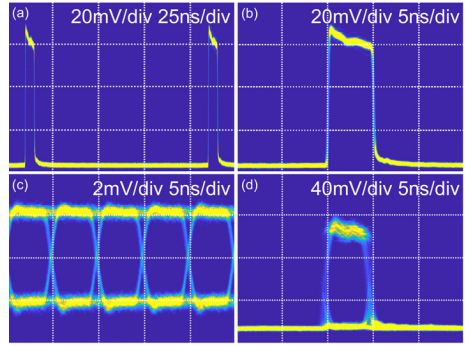

Fig. 4 Demonstration of optical sampling (a) Laser pulse sequence (b) Enlarged view of (a) that shows the width of a pulse (c) Eye pattern of the original signals without pulse sampling (d) Sampled signals by laser pulses.

Interference is introduced to demonstrate the effectiveness of the optical pulse sampling for BSS. Fig. 5 shows that the 2$^{nd}$ order moments measured from pulse-sampled signals match with the theoretical results. The theoretical results calculate the 2$^{nd}$ order moments with signals sampled at Nyquist frequency. In this experiment, the SOI $s_{soi}$ is random binary signal with data rate of 200Mbps, and the interference $s_{int}$ is Gaussian noise with bandwidth overlapped with the bandwidth of SOI and clock unsynchronized with SOI. The mixing matrix (Equation 1) used in the experiment is $a_{11} = a_{22} = 1$, and $a_{12} = a_{21} = 0.5$. An arbitrary waveform generator is used to generate the mixed signals $x_1$ and $x_2$. Fig. 5 (a) shows the $x_1$ and $x_2$ used in the experiment (black dots), and the theoretically calculated 2$^{nd}$ order moments (red curve). Each black dot is a sampled data point. The dots form the shape of two straight lines because the SOI is binary signal. Within each line, the dots are Gaussian distributed, which reflect the distribution of the interference. The six large black dots in Fig. 5 (b) show the experimentally measured 2$^{nd}$ order moments in six different angles ($\theta$ in Equation 3 and Fig. 2 (a)). The black curve is the fitting result based on the experimental measurement. The fitting curve matches with the theoretical result (red curve). Figs. 5 (c) and (d) show the time domain signals measured with an oscilloscope and are used to calculate 2$^{nd}$ order moments in Fig. 5 (b). Two points out of the six points are shown, and correspond to $x_1$, where $\theta = 0°$ (Fig. 5 (c)), and $x_2$, where $\theta = 90°$ (Fig. 5 (d)), respectively. In Fig. 5 (b), the 2$^{nd}$ order moment of $x_1$ is significantly smaller than the 2$^{nd}$ order moment of $x_2$, which is also clearly shown in Figs. 5 (c) and (d), where the variance of the signals in Fig. 5 (c) is significantly smaller than the variance of the signals in Fig. 5 (d).

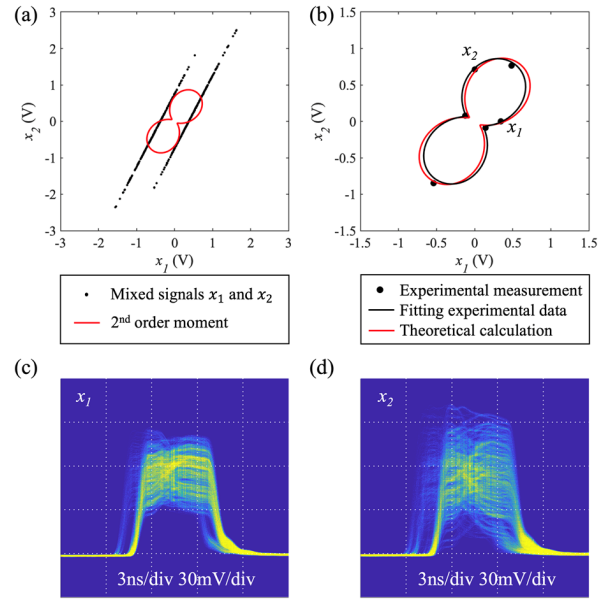

Fig. 5 Principal component analysis with signals sampled by optical pulses (a) Theoretical calculation of $2^{nd}$ order moments (red curve) and scatter plot for the mixed signals $x_1$ and $x_2$ (black dots) (b) Comparison between theoretical calculation (red curve) and experimental measurement of $2^{nd}$ order moments (black dots) (c) Experimentally measured histogram for $x_1$ (c) Experimentally measured histogram for $x_2$.

The de-mixing matrix is calculated with the pulse-sampled signals, and is applied to the mixed signals $x_1$ and $x_2$. Fig. 6 (a) shows that the recovered SOI matches with the original SOI, and a clear eye pattern is obtained Fig. 6 (b). The pulse width in Figs. 6 (a) and (b), or in another word, the duration of each sample is the same as the bit length of the original SOI. The bit error rate of the recovered signals is less than $1\times10^{-5}$. Figs. 6 (c) and (d) show that if the short optical pulse is not used, and the duration of each sample is five times the bit length of the SOI, the SOI is not separated with the Gaussian interference. The recovered SOI does not match with the original SOI (Fig. 6 (c)) and a noisy eye pattern is obtained (Fig. 6 (d)). The bit error rate is $9\times10^{-2}$, which is beyond the forward error correct limit [17].

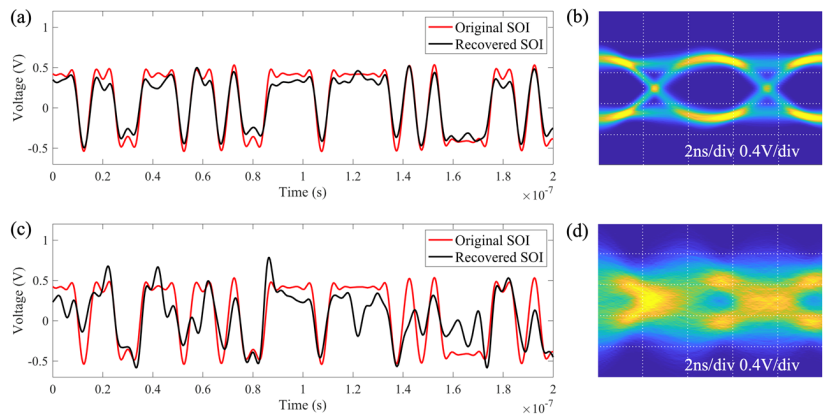

Fig. 6 Comparison of the recovered signals between short pulse sampling and long pulse sampling. The duration of each sample in (a) and (b) is the same as the SOI bit length. (a) shows the recovered SOI matches with the original SOI (b) shows a clear eye pattern. The duration of each sample in (c) and (d) is five times of the SOI bit length. (c) shows the recovered SOI does not match with the original SOI (d) shows a noisy eye pattern.

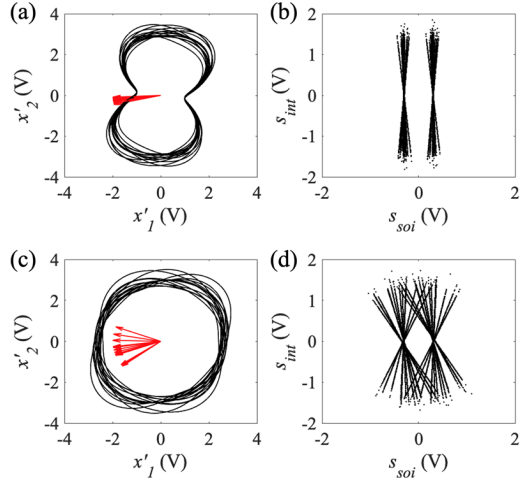

Fig. 7 Comparison of 4$^{th}$ order moments and the recovered signals between short pulse sampling and long pulse sampling. Both cases are tested with 15 groups of random signals. The duration of each sample in (a) and (b) is the same as the SOI bit length. (a) shows the 15 groups of random signals result in similar 4$^{th}$ order moments. (b) shows that all the 15 tests recover the SOI. The duration of each sample in (c) and (d) is five times of the SOI bit length. (c) shows the 15 groups of signals result in different and random 4$^{th}$ order moments. (d) shows that not all the 15 tests recover the SOI.

To explain the results in Fig. 6, we test the system with 15 groups of signals, and each group includes a random binary SOI and a random Gaussian interference (Fig. 7). $x'_1$ and $x'_2$ in Fig. 7 (a) are the mixed signals after PCA. Based on Equations 7 and 8, 4$^{th}$ order moments of $x'_1$ and $x'_2$ are calculated to perform ICA. The key parameter for ICA is $\phi_0$ in Equation 8. This is because in a scatter plot, matrix $V$ (Equation 9) corresponds to rotate the scatter plot by an angle $\phi_0$. If $\phi_0$ is applied in a right way, the projections of the two lines on horizontal axis of the scatter plot does not have any overlaps (Fig. 7 (b)), which means the SOI is completely separated from the interference. If $\phi_0$ is not applied in a right way, the projections of the two lines on horizontal axis of the scatter plot have overlaps (Fig. 7 (d)), which means the SOI is not completely separated from the interference. Figs. 7 (a) and (b) use short pulse sampling, where the duration of each sample is the same as the bit length of SOI. The $\phi_0$, which is marked by red arrows in Fig. 7 (a), from all the 15 groups of tests varies in a very small range (less than 10º). The projections of the rotated signals on the horizontal axis of the scatter plot does not have any overlaps (Fig. 7 (b)). Fig. 7 (b) plots the rotated signals for 15 tests together, and each parallel line corresponds to one test. All the lines in Fig. 7 (b) are close to each other with a variance of angle less than 10º. Figs. 7 (c) and (d) use long pulse sampling, where the duration of each sample is five times of the bit length of SOI. The $\phi_0$ (red arrow in Fig. 7 (c)) from 15 groups of tests varies in a large range (larger than 45º). The projections of the rotated signals on horizontal axis of the scatter plot have overlaps (Fig. 7 (d)). Such overlaps result in bit errors and a noisy eye pattern (Fig. 6 (d)).

### 3.2 Linearity analysis of mode-locked laser

Mode-locked laser generates femto-second laser pulses, which can be used to sample signals with 100GHz to THz bandwidth. Although femto-second pulses achieve the ultimate goal of ultra-fast sampling, the peak pulse power is orders of magnitude larger than its average power, and thus can saturate the photodetector of the BSS system (Fig. 1). By intensity modulating the

RF signals on the laser pulses, the pulse power has to be within the linear range of the photodetector to perform effective signal processing. Both the measurement of 2nd and 4th order moments calculation and interference cancellation require the system to operate in the linear range. More importantly, the linear range determines the capability of the system to separate strong interference from weak SOI. The upper bound of the linear range is defined by the saturation power of the photodetector, and the lower bound of the linear range is defined by the noise-equivalent power. This section studies the saturation effect of the photodetector and the linear range of the sampling system with a mode-locked laser. Experimental results show that the linear range is at least 30dB, which means that the system is able to separate interference that is 30dB stronger than the SOI.

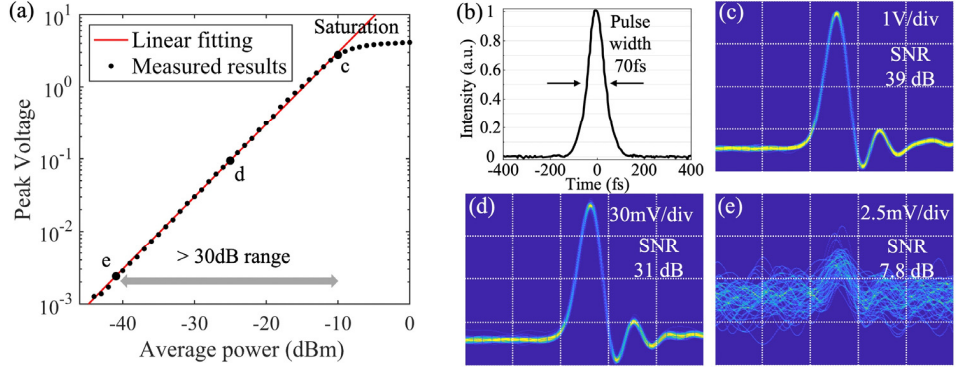

Fig. 8 (a) Peak voltage of the photodetector vs average power mode-locked laser, the small black dots are experimental results. (b) Femto-second pulse width measured with a correlator [18]. (c), (d) and (e) corresponds to points c, d, and e in (a), and are received pulses with average power of -10dBm, -25dBm, and -41dBm respectively. The time scales for (c), (d) and (e) are all 200ps/div.

The mode-locked laser used in this experiment has a pulse width of 70fs (Fig. 8 (b)), center wavelength of 1560nm, and a repetition rate of 37MHz (Precision Photonics FFL1560-MP [18]). The laser pulses are attenuated and then received with a photodetector with 50GHz bandwidth (Finisar XPDV21). The RF output of the photodetector is amplified by an RF amplifier with 9GHz bandwidth and then measured with an oscilloscope with 6GHz bandwidth (Keysight 6000X). Fig. 8 (a) shows the measured peak pulse voltage from the oscilloscope with the average power of laser scans from -45dBm to 0dBm. The average power is controlled by a digital attenuator. The measured peak pulse voltage increases linearly with the average power and saturates at -10dBm. The saturation power of a photon detection system to the pulse laser can be calculated as [19]:

$$P_{sp} = P_{scw} \frac{t_p}{t_{FWHM}} \qquad (10)$$

Where the $P_{sp}$ is the average power of the laser pulses that saturate the photodetector, $P_{scw}$ is the power the continuous wave laser that saturates the photodetector, $t_p$ is the pulse period, and the $t_{FWHM}$ is the full width at half maximum of the system impulse response. The photodetector used in this experiment has a $P_{scw}$ of 16dBm, the $t_{FWHM}$ is measured to be 90ps (Fig. 8 (c)). To measure the $t_{FWHM}$ of the system, a 70fs pulse from the mode-locked laser can be treated as an impulse (70fs << 90ps). With $P_{scw}$, $t_p$, and $t_{FWHM}$, $P_{sp}$ is calculated as -9dBm, which is close the measured saturation power -10dBm.

The SNR at -10dBm average power is 39dBm (Fig. 8 (c)). The signal is defined as the peak voltage and the noise is defined as the variance of the signals at the peak. The SNR drops with the average power. The SNR is 7.8dB when the average power is -41dBm, and the received histogram have observable pulse peak (Fig. 8 (d)). The SNR is close to 1dB when the average power is -44dBm. The pulse sampling system has a linear power range of over 30dB (-10dBm to -44dBm). The linear power range for the sampling system is important for BSS, since the separation of strong interference from weak SOI is a widely existing application scenario for the system. Such application requires the system to be able to address both the high and low power levels at the same time.

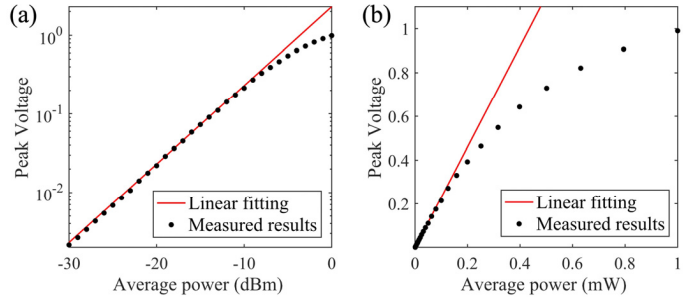

Fig. 9 Peak voltage of the received femto-second pulse vs average power without RF amplifier in (a) logarithmic scale and (b) linear scale

In addition to the 30dB range, we find that beyond the saturation power (-10dBm), the peak pulse voltage response of the photodetector still increases monotonously and significantly (Fig. 9). The results are repeatable, which means the photodetector does not break down. The significant increase of pulse voltage beyond -10dBm is not observed in Fig. 8 (a), because it reaches maximum output power of the RF amplifier. To remove the saturation effect of RF amplifier, Fig. 9 is measured without RF amplifier and shows that although the output response of the photodetector is no longer linear beyond the saturation power of -10dBm, the peak voltage still increases significantly with the input average power and the SNR is over 40dB. By calibrating the peak pulse voltage of the system, the nonlinear response can be corrected with the calibrated scale, and the -10dBm to 0dBm region can also be used to carry RF signals for BSS, which increase the dynamic range of the BSS system with another 10dB.

## 4. Conclusion

We proposed and experimentally demonstrated an optical pulse sampling system for photonic BSS. The photonic BSS separates the mixed signals in an analog way, and only statistical information of the mixed signals is needed to find the de-mixing matrix. The system is able to separate the SOI from interference at sampling rate orders of magnitude smaller than the data rate. The optical pulses sample the fast-changing signals for low-speed ADC without interference from adjacent bits. Experimental results show that with the pulse sampled signals, the statistical information can be accurately measured for BSS. By using femto-second pulses generated from mode-locked lasers, signals with bandwidth of 100GHz or more and power range of 30dB can be effectively sampled.

## Funding

This research is supported by Rowan University startup grant and New Jersey Health Foundation (Grant # PC 77-21).

## Acknowledgments

The authors would like to thank Yue-Kai Huang from NEC Laboratories America Inc to share the mode-locked laser for test and measurement in this paper.